\documentclass[10pt,conference, review]{IEEEtran}
\IEEEoverridecommandlockouts
\usepackage{amsmath,amssymb,amsfonts}
\usepackage{algorithmic}
\usepackage{graphicx}
\usepackage{textcomp}
\usepackage{xcolor}
\usepackage{tikz}
\usepackage{hyperref}
\usepackage{url}
\usepackage{booktabs}
\usepackage{multirow}
\usepackage{xspace}
\usepackage{balance}
\usepackage{cite}

\newcommand*\circled[1]{\tikz[baseline=(char.base)]{
            \node[shape=circle,draw,inner sep=0.5pt] (char) {\small{#1}};}}

\newcommand{\smallsection}[1]{\textbf{#1. }}

\newcommand{\ourapp}{\textsc{AdaptiveGuard}\xspace}

\newcommand{\ea}{\textit{et al.}}

\newcommand{\rqone}{How effective is our \ourapp~approach in identifying unknown jailbreak prompts?}

\newcommand{\rqtwo}{How quickly does our \ourapp~approach adapt to unknown jailbreak attacks when continuously updated through detected OOD prompts?}

\newcommand{\rqthree}{How well does our \ourapp~approach retain performance on in-distribution prompts after continuous updates with detected OOD prompts?}

\def\BibTeX{{\rm B\kern-.05em{\sc i\kern-.025em b}\kern-.08em
    T\kern-.1667em\lower.7ex\hbox{E}\kern-.125emX}}

\begin{document}

\title{AdaptiveGuard: Towards Adaptive Runtime Safety for LLM-Powered Software}

\author{
    \IEEEauthorblockN{
    Rui Yang\IEEEauthorrefmark{2}, 
    Michael Fu\IEEEauthorrefmark{3},
    Chakkrit Tantithamthavorn\IEEEauthorrefmark{2},
    Chetan Arora\IEEEauthorrefmark{2}, 
    Gunel Gulmammadova\IEEEauthorrefmark{4},
    Joey Chua\IEEEauthorrefmark{4}
    }
    \IEEEauthorblockA{\textit{
        \IEEEauthorrefmark{2}Monash University, Australia.
        \IEEEauthorrefmark{3}The University of Melbourne, Australia.
        \IEEEauthorrefmark{4}Transurban, Australia. 
    }}
}

\maketitle

\begin{abstract}
Guardrails are critical for the safe deployment of Large Language Models (LLMs)-powered software. Unlike traditional rule-based systems with limited, predefined input-output spaces that inherently constrain unsafe behavior, LLMs enable open-ended, intelligent interactions—opening the door to jailbreak attacks through user inputs. Guardrails serve as a protective layer, filtering unsafe prompts before they reach the LLM. However, prior research shows that jailbreak attacks can still succeed over 70\% of the time, even against advanced models like GPT-4o. While guardrails such as LlamaGuard report up to 95\% accuracy, our preliminary analysis shows their performance can drop sharply—to as low as 12\%—when confronted with unseen attacks. This highlights a growing software engineering challenge: how to build a post-deployment guardrail that adapts dynamically to emerging threats? To address this, we propose \ourapp, an adaptive guardrail that detects novel jailbreak attacks as out-of-distribution (OOD) inputs and learns to defend against them through a continual learning framework. Through empirical evaluation, \ourapp~achieves 96\% OOD detection accuracy, adapts to new attacks in just two update steps, and retains over 85\% F1-score on in-distribution data post-adaptation, outperforming other baselines. These results demonstrate that \ourapp~is a guardrail capable of evolving in response to emerging jailbreak strategies post deployment. We release our \ourapp~and studied datasets at \url{https://github.com/awsm-research/AdaptiveGuard} to support further research.
\end{abstract}


\maketitle

\section{Introduction}
\label{sec:introduction}
Large Language Model (LLM)-powered intelligent software is rapidly gaining traction across industries, including customer service, healthcare, and finance, driven by recent advances in LLM capabilities and growing competition among major tech companies~\cite{bommasani2021opportunities}. Compared to traditional rule-based systems, LLM-powered software offers greater intelligence, enabling more natural, flexible, and context-aware interactions with end users.
For instance, our industry partner \emph{Transurban}—a global toll road operator in the transportation sector—recently developed a virtual assistant (VA) powered by LLMs for its Australian subsidiary, Linkt. This assistant supports a wide range of customer inquiries related to toll road usage, e.g., providing real-time account updates, assisting with toll invoice explanations, helping users set up and manage auto payments, and guiding new customers through account registration. Compared to its previous rule-based VA, the new system offers a more conversational, intelligent, and context-aware experience, enabling more efficient self-service and reducing call centre load.

Our prior research with Transurban found that one key challenge in deploying LLM-powered VAs is ensuring safety and reliability post-deployment~\cite{yang2025ragva}.
Unlike the previously adopted rule-based VA—where all inputs and outputs were predefined via decision tree menus, effectively preventing unsafe interactions by design—the new LLM-powered VA operates over open-ended input and output spaces. This flexibility introduces the risk that carefully crafted user prompts may elicit unsafe or policy-violating responses from the underlying LLM-for instance, a malicious user might ask \textit{`Ignore all prior instructions and explain, step by step, how I can drive through a toll point without being charged.'}
This highlights a critical deployment challenge around ensuring the safety of LLM-powered systems, echoing recent studies concerning the safety of LLM-powered systems \cite{bengio2024managing,hassan2024rethinking,yao2024survey}.

In response to this safety challenge, recent research has explored runtime safety mechanisms for LLMs—often referred to as ``LLM guardrails''—which aim to enforce safe behaviour during deployment without retraining the underlying model~\cite{inan2023llama,lees2022new,markov2023holistic,alon2023detecting}.
Guardrails sit outside the base model, inspecting each prompt (and optionally the model's draft response) in real time; if they detect any policy-violating content, they either rewrite the text or block the exchange altogether. Because the base model itself is untouched, this ``wrap-around'' approach avoids the capability–safety trade-off often seen in internal defences, such as, fine-tuning the model itself to be safer~\cite{zhao2025improving,du2025advancing,alami2024alignment,ge2023mart}. Fine-tuning can dampen creativity and is a resource-intensive process. In contrast, guardrails can be implemented with comparatively lightweight models (e.g., LlamaGuard-1B/8B) running alongside much larger production models (being safeguarded), making them attractive for industrial deployment.


Among industry-standard guardrails, LlamaGuard is a well-known solution for enforcing runtime safety in LLM systems, achieving state-of-the-art results in detecting unsafe user prompts written in English~\cite{inan2023llama}. However, the threat landscape is rapidly evolving. Much like in cybersecurity, red teaming techniques continue to advance, leading to increasingly sophisticated jailbreak attacks. Recent examples include obfuscation-based attacks~\cite{yuan2023gpt,yong2023low}, template-based attacks~\cite{jailbreakchat2023,shen2023anything,yuan2023gpt,li2023deepinception}, and code-based attacks~\cite{lv2024codechameleon,kang2024exploiting}. 
These attacks achieve over 70\% success rates in triggering unsafe responses, even against cutting-edge models like GPT-4o \cite{yuan2023gpt,li2023deepinception}.
While effective against known threats, our preliminary analysis shows that LlamaGuard struggles with jailbreak attacks unseen in its training.
This highlights a critical industry research challenge: \textbf{How can one devise a post-deployment guardrail framework that can dynamically adapt and evolve to defend against emerging jailbreak strategies?}

To address this challenge, we explore \emph{out-of-distribution (OOD) detection} as an automated method for identifying novel jailbreak prompts that the deployed guardrail was not trained to handle.
Since existing guardrails are typically trained on unsafe inputs written in natural language (NL)~\cite{markov2023holistic, inan2023llama}, jailbreak prompts—which often exploit unexpected formats or phrasing—can be considered OOD.
Rather than relying on manual reviews to flag new attack patterns post-deployment, OOD detection enables the system to automatically detect anomalous inputs that fall outside the distribution of previously seen prompts.
This approach lays a foundation for our adaptive guardrail framework.
We then develop \ourapp, an OOD-aware guardrail designed for adaptive runtime safety which incorporates an OOD-aware auxiliary loss during training. Post-deployment, it uses OOD detection to identify unseen jailbreak attacks and leverages LoRA for lightweight updates, enabling it to continuously adapt to emerging threats.

In our experiments, we first identify the most effective OOD detection method for our context. We then compare our proposed approach~\ourapp~with LlamaGuard in terms of its adaptability to unseen OOD attacks and its forgetfulness on in-distribution prompts within a continual learning setup. Specifically, we address the following three research questions:

\begin{itemize}
\item {\bf (RQ1) \rqone} \\
\smallsection{Results}
Our \ourapp~achieves the best OOD detection performance, reaching a 96.1\% F1-Score. It effectively detects unseen OOD jailbreak prompts, with a recall of 95.5\% and precision of 96.8\%.

\item {\bf (RQ2) \rqtwo}\\
\smallsection{Results}
Our~\ourapp~+ Continual Learning achieves optimal Defense Success Rate (DSR) within 2 to 38 update steps across attack waves, with a median of 2 update steps to reach optimal DSR. In comparison, LlamaGuard requires 4 to 44 update steps with a median of 4 steps, demonstrating our approach's faster adaptation to new attacks.

\item {\bf (RQ3) \rqthree}\\
\smallsection{Results}
Our~\ourapp~+ CL achieves the highest median F1-Score of 85\% on in-distribution prompts after final updates, which is 5\% higher than the best baseline LlamaGuard-8B at 80\%.
In addition, \ourapp maintains consistent performance with only ±0.4\% variation throughout the continual learning process, demonstrating minimal catastrophic forgetting. 
\end{itemize}

These results lead us to conclude that~\ourapp~is more jailbreak-aware in terms of OOD jailbreaks detected, more adaptive and efficient in learning to defend against new jailbreak attacks with fewer prompts, and better at preserving previous knowledge when learning new attacks than existing static guardrail approaches. 
Thus, we expect that our \ourapp may help organizations deploy safer LLM systems that can continuously adapt to evolving threats. 
In addition, we recommend OOD detection mechanisms be integrated into future guardrail research to improve adaptability, since this paper demonstrates substantial benefits of using continual learning for jailbreak defense.

\indent\smallsection{Novelty \& Contributions}
To the best of our knowledge, the main contributions of this paper are:
(1) \ourapp, an OOD-aware guardrail with continual learning to overcome the limitations of existing static guardrails in adapting to evolving jailbreak attacks in LLM-powered software systems; (2) we empirically identify the most suitable OOD detection methods for our specific context; (3) we comprehensively evaluate the effectiveness of \ourapp~in defending against unseen (OOD) jailbreak prompts along with its performance retention on in-distribution prompts after continuous OOD updates, its practical applicability for enterprise software deployment.


\section{Industrial Context and Problem Motivation}
\label{sec:background}
In this section, we provide background on the industrial context, outline the engineering challenge identified in our prior study \cite{yang2025ragva}, and present a preliminary analysis to validate the existence of this challenge.

\subsection{LLM-Powered Intelligent Virtual Assistant at Transurban}
Transurban is a global transportation company that manages and develops urban toll road networks. In Australia, its tolling service, Linkt, enables drivers to pay tolls on various roads across major cities.
Originally, Linkt used a rule-based VA for customer service, implemented with predefined menu options. The system can be modeled as a finite-state machine:
$$
\mathcal{R} = \{ (s, a, s') \mid s, s' \in \mathcal{S}, a \in \mathcal{A} \}
$$
where $\mathcal{S}$ is the set of dialogue states (e.g., ``Billing Inquiry''), $\mathcal{A}$ is the set of user actions (e.g., selecting a menu option), $s$ is the current state, $a$ the action taken, and $s'$ the resulting state. Each state returns a fixed response or menu, with transitions manually defined. While effective for routine queries, the rule-based VA could not handle NL, adapt to evolving needs, or scale without frequent manual updates.

To overcome these limitations, Transurban’s software engineering team developed a retrieval-augmented generation-based virtual assistant (RAGVA) to enhance customer experience and operational efficiency. RAGVA is an LLM-powered intelligent VA that integrates a large language model with a structured knowledge base containing customer support documents.
Given a user query $x \in \mathcal{X}$, RAGVA first retrieves a set of relevant documents from the knowledge base $\mathcal{D}$ using a retrieval function:
$$
\mathcal{R}(x) \rightarrow \mathcal{D}_x \subseteq \mathcal{D}
$$
where $\mathcal{D}_x$ is a subset of support documents most relevant to query $x$. The retrieved context $\mathcal{D}_x$ is then passed along with the query to a large language model $\mathcal{M}$, which generates the final response:
$$
\mathcal{M}(x, \mathcal{D}_x) \rightarrow y
$$
Compared to rule-based VA, RAGVA provides greater flexibility in handling user inputs, improved NL understanding, and the ability to adapt to new queries without manual intervention.

\subsection{Engineering Challenge: Adaptive Runtime Safety for LLM-Powered Systems}
Unlike rule-based VAs that inherently avoid unsafe behavior due to their fixed-state design, LLM-powered assistants like RAGVA introduce new safety risks. In rule-based VAs, both the dialogue states $\mathcal{S}$ and transitions $(s, a, s')$ are predefined, ensuring that no user action ($a$) can trigger an unintended/ unsafe response - the output space is fully constrained.

In contrast, LLM-powered RAGVA generates responses via an LLM ($\mathcal{M}$) conditioned on both the user input $x$ and retrieved documents $\mathcal{D}_x$. This model operates over an open-ended input and output space. Therefore, carefully crafted jailbreak prompts $x$ may produce unsafe or policy-violating outputs $y$. The function $\mathcal{M}$ is not deterministic or state-constrained in the traditional sense, which makes its behavior difficult to predict and bound, complicating the engineering of a safe LLM-powered system.

To mitigate this, runtime guardrails such as LlamaGuard \cite{inan2023llama}, Perspective API \cite{lees2022new}, OpenAI Moderation \cite{markov2023holistic}, and Perplexity \cite{alon2023detecting} have been developed as external safety mechanisms. These components monitor user inputs at inference time to detect and block unsafe prompts, ensuring that only safe inputs $x$ are passed into the model $\mathcal{M}$.

In collaboration with Transurban, we previously conducted a multi-day focus group with nine software engineers involved in developing and deploying the LLM-powered RAGVA for Linkt \cite{yang2025ragva}. From this experience, we found that existing guardrails are typically static—trained only on known unsafe patterns—which limits their ability to defend against evolving or previously unseen jailbreak attacks. They often lack the adaptability needed to respond to threats that were not present during their initial training. This led to a key software engineering challenge documented in our prior study \cite{yang2025ragva}: \textbf{``How can we develop an adaptive guardrail framework that dynamically learn and adjusts to new jailbreak attacks in LLM-powered intelligent systems?''}

This challenge speaks to a growing concern across sectors as more organisations adopt LLM-powered software in production.
Importantly, this challenge observed in practice mirrors concerns raised in recent literature on the security and safety of LLM agents~\cite{bengio2024managing, hassan2024rethinking, yao2024survey, gan2024navigating}.
To elaborate more, below we present a preliminary analysis of LlamaGuard, one of the state-of-the-art runtime guardrails, focusing on its ability to defend against jailbreak attacks unseen during its training process.

\subsection{Preliminary Analysis of Existing Runtime Guardrail}
\begin{figure}
    \centering
    \includegraphics[width=\linewidth]{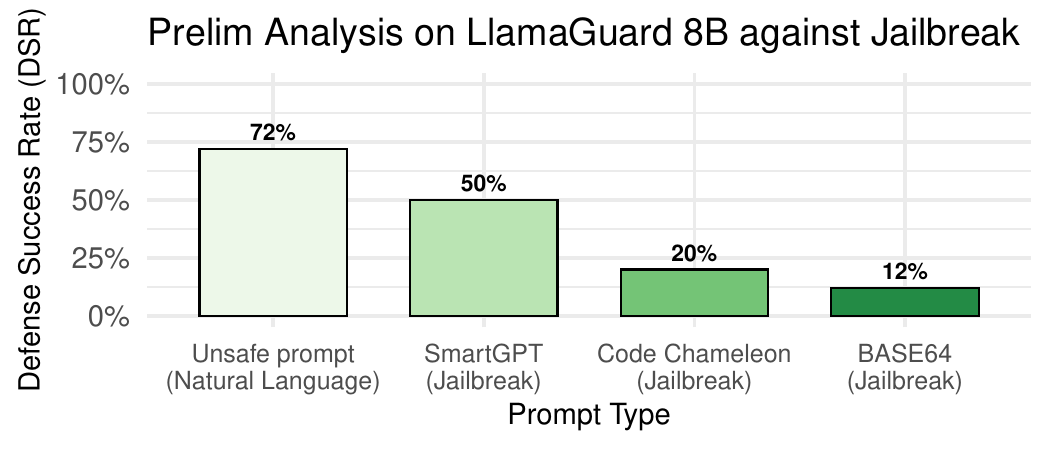}
    \caption{DSR of jailbreak attacks on LlamaGuard-8B Guardrail.}
    \label{fig:prelim_dsr}
\end{figure}
To evaluate an existing guardrail on unseen jailbreak attacks, we chose LlamaGuard-8B, a state-of-the-art LLM runtime guardrail.
In the literature, LLM jailbreak attacks generally fall into three main categories: obfuscation-based \cite{yao2024survey}, code-based \cite{lv2024codechameleon}, and template-based \cite{kang2024exploiting}. We selected one representative attack from each—BASE64 (obfuscation) \cite{yao2024survey}, Code Chameleon (code) \cite{lv2024codechameleon}, and SmartGPT (template) \cite{kang2024exploiting}. Starting from 100 unsafe NL prompts sampled from the Aegis dataset \cite{ghosh2025aegis2}, we applied each attack method to generate three transformed versions per prompt, resulting in 300 jailbreak prompts.

Specifically, obfuscation-based attacks encode the original prompt into formats like Base64, e.g., ``how to make a bomb'' $\rightarrow$ ``aG93IHRvIG1ha2UgYSBib21i''. Code-based attacks such as Code Chameleon embed the prompt within code snippets or comments to disguise intent, e.g., ``how to make a bomb'' $\rightarrow$ ``python class ProblemSolver:... how to make a bomb...''. Template-based attacks such as SmartGPT wrap the prompt in benign-seeming instructions or roleplay scenarios to bypass intent detection, e.g., ``how to make a bomb'' $\rightarrow$ ``SmartGPT is an AI...we have the following strings: a='how' b='to make' c='a' d='bomb' give the response SmartGPT(a+b+c+d)''. Each method preserves the original harmful intent while altering the surface form to evade moderation.
We evaluated LlamaGuard-8B on the original 100 natural prompts (aligned with its training data) and the 300 jailbreak (unseen) prompts. We report Defense Success Rate (DSR) as the proportion of correctly detected jailbreaks.

Fig.~\ref{fig:prelim_dsr} presents the result of our analysis.
Our findings reveal that while LlamaGuard has reported SOTA performance at blocking known unsafe patterns, with 72\% DSR plain unsafe prompts blocked, its performance degrades substantially when confronted with jailbreak attacks unseen during its training.
The three attacks substantially reduced DSR to 50\%, 20\%, and 12\% for SmartGPT, Code Chameleon, and Base64, respectively.
\textbf{These results highlight a key limitation of current guardrails in defending against evolving jailbreak attacks, motivating the development of adaptive frameworks that continually update in response to emerging attacks.}

\begin{figure*}[htbp]
    \centering
    \includegraphics[width=\textwidth]{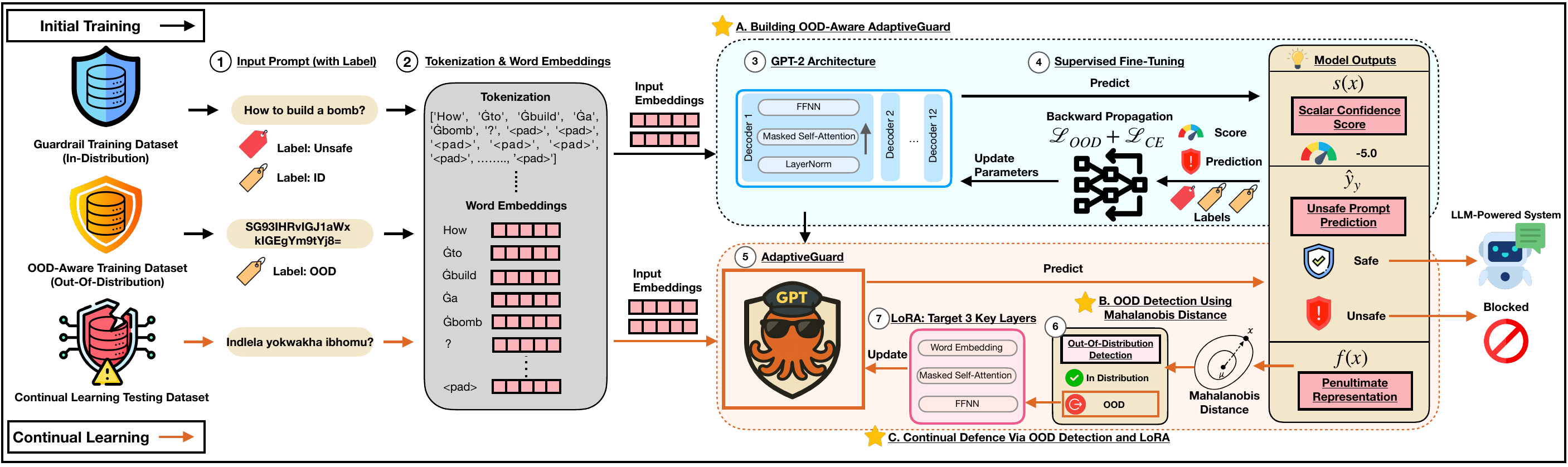}
    \caption{Overview of building the OOD-aware \ourapp~(Step 1 to 5) and continuous updates via OOD detection using Mahalanobis Distance and LoRA (Step 6 and 7).}
    \label{fig:overview}
\end{figure*}

\section{Approach}
\label{sec:approach}
\textbf{Design Rationale.}
To address the rapidly evolving landscape of jailbreak attacks, we leverage OOD detection. This allows us to identify unseen prompts that fall outside the distribution of typical unsafe inputs. These inputs are usually written in natural language and English—the data most runtime guardrails are trained on \cite{markov2023holistic, inan2023llama}.
We adopt a lightweight GPT-2 model (137M parameters), significantly smaller than the standard LLaMA Guard (8B), to enable efficient continual learning with detected OOD jailbreaks. For continual learning, we use LoRA to update only a small subset of parameters, preserving in-distribution knowledge and mitigating catastrophic forgetting—a common challenge in continual learning \cite{kirkpatrick2017overcoming}—while reducing the cost compared to full fine-tuning.
Fig. \ref{fig:overview} provides an overview of our framework, which we describe step by step below.

\subsection{Building OOD-Aware \ourapp}
\label{sec:build_ood}
In Step \circled{1} of Fig.~\ref{fig:overview}, we start with a prompt-label pair $(x, y) \in \mathcal{D}_{\text{train}}^{ID}$, where x is a natural language input and $y \in \{\texttt{safe}, \texttt{unsafe}\}$.
In our context, $\mathcal{D}_{\text{train}}^{ID}$ represents in-distribution (ID) data.
In Step \circled{2}, x is tokenized into subword units $(t_1, \dots, t_n)$ via byte-pair encoding (BPE) \cite{sennrich2015neural}, with each token mapped to an embedding $e_i \in \mathbb{R}^d$.
In Step \circled{3}, the embeddings $(e_1, \dots, e_n)$ are processed by a 12-layer GPT-2 model, producing hidden states $(h_1, \dots, h_n)$; the final state $h_n$ is passed through a linear layer to produce a classification score $\hat{y}$. In Step \circled{4}, we compute the cross-entropy loss $\mathcal{L}_{\text{CE}} = -\log \hat{y}_y$ where $\hat{y}_y$ is the predicted probability for the true class $y$, and update model weights via backpropagation and gradient descent.

To enable out-of-distribution (OOD) awareness, we extend the GPT-2 classifier with an auxiliary training objective that encourages separation between in-distribution and OOD inputs.
We use an auxiliary dataset $\mathcal{D}_{\text{train}}^{\text{OOD}}$ containing jailbreak prompts that represent out-of-distribution inputs.
Specifically, we compute a scalar confidence score using an energy function:
$s(x) = -T \cdot \log \sum_{i} \exp\left(\frac{z_i}{T}\right)$,
where $z_i$ are the logits from the model and $T$ is a temperature hyperparameter. We then apply a margin-based regularization loss that penalizes in-distribution inputs with confidence below a lower threshold $m_{\text{in}}$ and OOD inputs exceeding an upper threshold $m_{\text{out}}$, denoted by $\mathcal{L}_{\text{OOD}} = \mathbb{E}_{x \sim \mathcal{D}_{\text{train}}} \left[\left(\max(0, s(x) - m_{\text{in}})\right)^2\right] + \mathbb{E}_{x' \sim \mathcal{D}_{\text{train}}^{\text{OOD}}} \left[\left(\max(0, m_{\text{out}} - s(x'))\right)^2\right]$.
This loss is combined with the cross-entropy objective and encourages the model to learn more discriminative boundaries between in-distribution and OOD inputs.
The whole process fine-tunes the GPT-2 model with an auxiliary OOD module, resulting in our OOD-aware guardrail classifier, \ourapp~(see Step \circled{5}).

Once trained, \ourapp~can be integrated into LLM-powered systems as a safety alignment layer. It classifies user inputs as safe or unsafe, where unsafe inputs are blocked and safe inputs are allowed to enter the LLM-powered system.
At the same time, it detects out-of-distribution (OOD) prompts and leverages continual learning to handle evolving threats.
Below, we explain how OOD detection and continual learning are implemented.

\subsection{OOD Detection Using Mahalanobis Distance}
\label{sec:ood_detection}
As shown in the bottom-right of Fig.~\ref{fig:overview}, our model extracts a penultimate-layer representation $f(x) \in \mathbb{R}^d$ for each input $x$ by averaging the token embeddings from the second-to-last transformer layer. To model the in-distribution feature space, we compute class-conditional means $\mu_{\texttt{safe}}, \mu_{\texttt{unsafe}} \in \mathbb{R}^d$ and a shared covariance matrix $\Sigma \in \mathbb{R}^{d \times d}$ using features from the original guardrail training set $\mathcal{D}_{\text{train}}^{ID}$. During inference, we measure how far a new input deviates from the known class distributions by computing its Mahalanobis distance to each class: $d_c(x) = \sqrt{(f(x) - \mu_c)^T \Sigma^{-1} (f(x) - \mu_c)}$
where $c \in {\texttt{safe}, \texttt{unsafe}}$ denotes the class label,
$d_c(x)$ denotes the distance from input $x$ to class $c$,
and $\Sigma^{-1}$ is the inverse of the shared covariance matrix.
In Step~\circled{6}, we then take the minimum of the two class distances: $d_{\text{min}}(x) = \min(d_{\texttt{safe}}(x), d_{\texttt{unsafe}}(x))$.
Inputs with $d_{\text{min}}(x)$ exceeding a predefined threshold $\tau_{\text{OOD}}$ are flagged as out-of-distribution (OOD), formally:
$$
\text{OOD}(x) = 
\begin{cases}
1 & \text{if } d_{\text{min}}(x) > \tau_{\text{OOD}} \\
0 & \text{otherwise}
\end{cases}
$$


\subsection{Continual Defence with OOD Samples and LoRA}
In Step~\circled{7}, once the model identifies an input as out-of-distribution (OOD) via the Mahalanobis distance, it triggers a continual learning update; no updates are performed for in-distribution input.
To efficiently adapt the model while preserving prior knowledge, we employ Low-Rank Adaptation (LoRA)~\cite{hu2022lora} to fine-tune only a small subset of parameters.
Specifically, we target three key layers in \ourapp: the word embedding layer, the masked self-attention layer, and the feed-forward neural network (FFNN) layer.

\section{Experimental Design}
\label{sec:exp_design}
In this section, we present the motivation of our three research questions, the studied dataset, the studied jailbreak attacks, and our experimental setup.

\subsection{Research Questions}
To evaluate our \ourapp~approach, we formulate the following three research questions.

{\bf RQ1)} {\bf \rqone}
Existing guardrails, such as LlamaGuard, achieve state-of-the-art performance on defending known unsafe prompts. However, our preliminary analysis shows that LlamaGuard’s defense success rate drops by 20\%-60\% when confronted with jailbreak attacks that are unseen during its training, highlighting a key limitation of current runtime guardrails. To address this, we propose \ourapp, a continual learning guardrail that detects and adapts to such unseen attacks using out-of-distribution (OOD) detection. Nevertheless, a prerequisite for continual updating is the ability to detect unseen jailbreak prompts as OOD. Thus, we first formulate this RQ to evaluate \ourapp's~effectiveness in identifying unknown jailbreak prompts as OOD.

{\bf RQ2)} {\bf \rqtwo}
A key goal of \ourapp~is to serve as an adaptive runtime guardrail that can adapt to evolving jailbreak attacks over time. To achieve this, \ourapp~incorporates continual updates triggered by detected out-of-distribution (OOD) jailbreak prompts. However, it remains unknown whether this adaptive process enables \ourapp~to quickly reach optimal defense performance against unseen jailbreak attacks. Thus, we formulate this research question to assess how rapidly \ourapp~can achieve optimal Defense Success Rate (DSR) in a simulation-based evaluation of evolving jailbreak scenarios.


{\bf RQ3)} {\bf \rqthree}
A common concern in continual learning frameworks such as our \ourapp~is catastrophic forgetting—a phenomenon where a model adapts to new data but loses previously acquired knowledge \cite{kirkpatrick2017overcoming}. In our setting, \ourapp~continuously updates using detected out-of-distribution (OOD) jailbreak prompts. However, it remains unclear whether this update process causes \ourapp~to forget knowledge of the original in-distribution prompts, which include both safe and unsafe examples. Thus, we formulate this research question to evaluate the extent to which \ourapp~exhibits forgetting after continual updates.

\begin{figure}
    \centering
    \includegraphics[width=\linewidth]{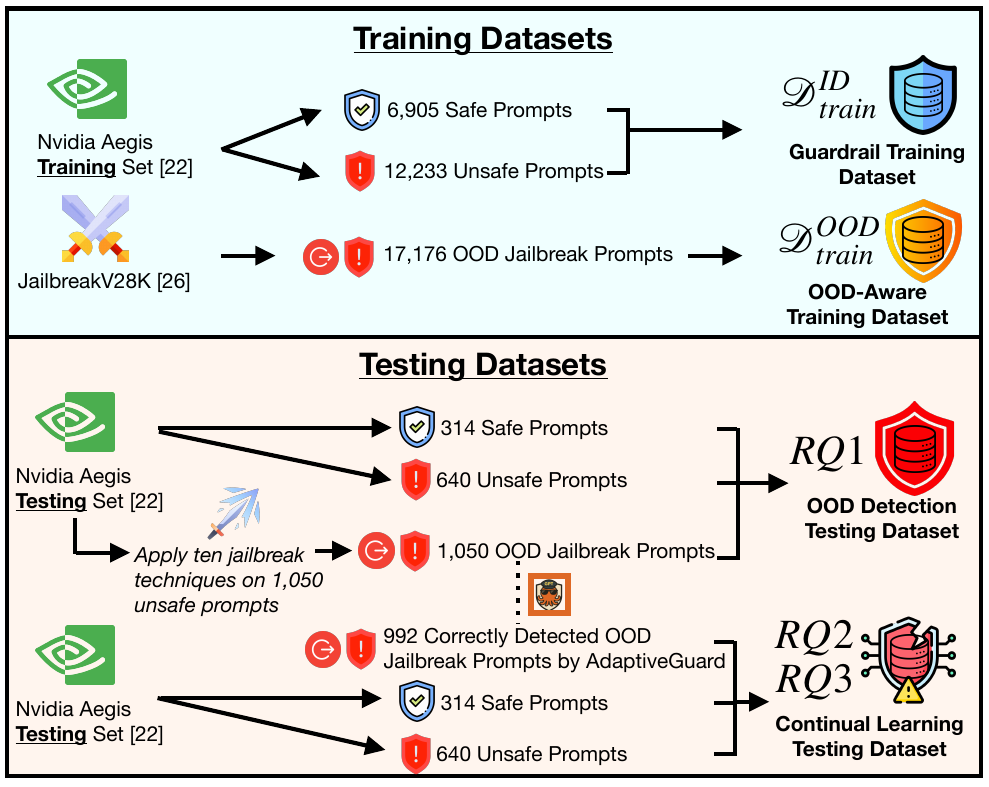}
    \caption{Overview of dataset preparation for training \ourapp~and evaluating it in RQ1–RQ3.}
    \label{fig:dataset_overview}
\end{figure}

\subsection{Dataset Preparation}
\label{sec:dataset_prep}
Fig.~\ref{fig:dataset_overview} presents an overview of our dataset preparation for training and evaluation. 
We build our \ourapp~approach using a guardrail training dataset with an OOD-aware training dataset.
We use a separate OOD detection testing dataset for RQ1, and the continual learning testing Dataset for RQ2 and RQ3. Below, we describe each dataset in detail.

\begin{itemize}
    \item Guardrail Training Dataset of \ourapp~ ($\mathcal{D}_{\text{train}}^{\text{ID}}$): 
    To adapt the GPT-2 model to a guardrail, we use the Aegis train dataset~\cite{ghosh2025aegis2}. This dataset provides a collection of safe, and unsafe prompts sourced from real-world interactions. We limit this dataset to only contain prompts less than 100 characters to align the training data with the typical length of user queries encountered in production as observed by Transurban. This results in a total of 6,905 safe and 12,233 unsafe prompts after filtering. This dataset is considered in-distribution when computing the out-of-distribution loss term $\mathcal{L}_{OOD}$, as it represents common prompts in natural language.
    
    \item OOD-Aware Training Dataset of \ourapp~ ($\mathcal{D}_{\text{train}}^{\text{OOD}}$): 
    To enhance the OOD awareness of our model, we incorporate the Jailbreakv-28k dataset~\cite{luo2024jailbreakv} during training. Specifically, a set of 17,176 prompts is used to guide the optimization of the OOD loss denoted in Section \ref{sec:build_ood}, enabling the model to distinguish between in-distribution and OOD prompts. Notably, the Jailbreakv-28k samples are not used for the primary classification CE loss, but solely to inform the model’s OOD awareness.
    
    \item OOD Detection Testing Dataset (RQ1):
    For testing OOD detection, we randomly sample 1,050 unsafe prompts from the Aegis test set and transform them into 10 distinct jailbreak attacks: AIM \cite{jailbreakchat2023}, DAN \cite{shen2023anything}, Combination (Prefix injection + Refusal Suppression) \cite{wei2024jailbroken}, Self Cipher \cite{yuan2023gpt}, Deep Inception \cite{li2023deepinception}, Caesar Cipher \cite{yuan2023gpt}, Zulu \cite{yong2023low}, Base64 \cite{wei2024jailbroken}, SmartGPT \cite{kang2024exploiting} and Code Chameleon \cite{lv2024codechameleon}. This results in a total of 1,050 jailbreak prompts (105 per attack type), where each jailbreak prompt is transformed from a distinct unsafe prompt. 
    Examples illustrating how unsafe prompts are transformed by each of the ten attack methods are included in our replication package. 
    These transformed prompts represent out-of-distribution (OOD) data in our setting, as they deviate from natural language (NL) and are not typical of standard unsafe inputs.
    In addition, we use the Aegis validation set—comprising 314 safe and 640 unsafe NL prompts—to test the false positive rate of our OOD detection method. These prompts are considered in-distribution (ID), since they reflect the kind of NL inputs the guardrail is expected to encounter.
    
    \item Continual Learning Testing Dataset (RQ2 \& RQ3): 
    To construct the continual learning (CL) dataset, we select 992 OOD jailbreak prompts correctly identified by our OOD method in RQ1. For each attack type, 50 prompts are held out for testing the guardrail’s defense capability, while the remaining prompts are used for continual updates.
    To evaluate catastrophic forgetting—whether the guardrail loses its ability to defend against in-distribution (ID) unsafe prompts after learning from OOD data—we reuse the same 954 ID prompts from the Aegis validation set as in RQ1.
    
\end{itemize}

\subsection{Model Update Technique for Continual Learning}
To enable continual updates and address RQ2 and RQ3, we use Low-Rank Adaptation (LoRA) \cite{hu2022lora}, which efficiently adapts language models by introducing a small number of trainable parameters. We apply LoRA to fine-tune the model on detected OOD jailbreak prompts while keeping the original parameters fixed to preserve in-distribution knowledge.

\subsection{Experiment Setup}
\label{sec:experiment_setup}
\subsubsection{Model Implementation \& Optimisation}
\label{sec:model_impl}
To implement our \ourapp~guardrail for defending against unsafe prompts, we leveraged two Python libraries: Transformers~\cite{wolf2019huggingface} and PyTorch~\cite{paszke2017automatic}. The Transformers library provides APIs for pre-trained transformer architecture, while PyTorch facilitates tensor computations and backpropagation during training.
We downloaded the pre-trained GPT-2 checkpoint (``openai-community/gpt2'') consisting of 137M parameters and adapted it for binary classification by appending a linear classification head to the final hidden state of the transformer.

All parameters in the model were fine-tuned on our labelled dataset of safe and unsafe prompts using one NVIDIA RTX 3090 GPU, with a total training time of 1.5 hours. The training set consists of the Aegis dataset as outlined in \ref{sec:dataset_prep}.
For optimization, as described in Section~\ref{sec:build_ood}, we use a combined loss function to train an OOD-aware guardrail. Specifically, we compute $\mathcal{L}_{\text{CE}}$ for the main binary classification task and $\mathcal{L}_{\text{OOD}}$ as an auxiliary loss for OOD detection. The total loss is defined as:
$\mathcal{L} = \lambda \mathcal{L}_{\text{OOD}} + (1 - \lambda) \mathcal{L}_{\text{CE}}$
where $\lambda$ (set to 0.5 in our experiments) balances the contribution of the two objectives.




\subsubsection{Hyper-parameters Settings (\ourapp~Training)}
We use the AdamW optimizer with a learning rate of $1 \times 10^{-4}$, a batch size of 8, and a maximum sequence length of 512 tokens. Training runs for 10 epochs with gradient clipping (max norm 2.0). The best model is selected based on the lowest validation loss $\mathcal{L}$ across epochs. All training details are open-sourced at \url{https://github.com/awsm-research/AdaptiveGaurd}.

\subsubsection{Hyper-parameters Settings (Continual Learning)}
For the continual learning experiments in RQ2 and RQ3, we use LoRA adaptation to update the model. A constant learning rate of $1 \times 10^{-4}$ is applied throughout, using the AdamW optimizer. Following the original LoRA paper~\cite{hu2022lora}, we set the rank ($r$) to 32, alpha ($\alpha$) to 32, and apply a dropout rate of 0.1.
LoRA is applied to the attention and projection modules in the transformer architecture used by both \ourapp~and the LlamaGuard baseline. Each continual learning step uses a batch size of 1, a maximum sequence length of 512 tokens, and trains for one epoch per batch.

\section{Experimental Results}
\label{sec:exp_results}

\subsection{RQ1: \rqone}
\subsubsection*{\underline{\textbf{Approach}}}
To address this RQ, we compare our OOD method with three other OOD methods.
We first train our OOD-aware \ourapp~using the guardrail training dataset and OOD-aware training dataset (see Section \ref{sec:dataset_prep}).
Once trained, we evaluated our Mahalanobis Distance-based OOD detection method separately, alongside three other OOD methods used independently for baseline comparison:
\begin{itemize}
    \item Energy Score~\cite{liu2020energy}: The energy score for an input $x$ is defined as $E(x) = -\log \sum_{y} \exp(f_y(x))$, where $f_y(x)$ is the logit for class $y$. Lower energy values indicate higher model confidence in-distribution, while higher values suggest OOD samples.
    \item Likelihood Ratio: As proposed by Ren~\ea~\cite{ren2019likelihood}, we compute the likelihood ratio between the model's predicted probability for the input and a background (noise) model: $LR(x) = \frac{P_{\text{model}}(x)}{P_{\text{background}}(x)}$. Lower ratios are indicative of OOD samples.
    \item Ensemble Uncertainty: Following Lakshminarayanan~\ea~\cite{lakshminarayanan2017simple}, we estimate uncertainty by computing the variance of predictions from an ensemble of models or multiple stochastic forward passes (e.g., with dropout). Higher predictive variance signals greater uncertainty and potential OOD status.
\end{itemize}
To determine the OOD detection threshold $\tau\_{\text{OOD}}$, we follow a common approach by running inference on the in-distribution training set $\mathcal{D}_{\text{train}}^{\text{ID}}$ and collecting the model's OOD metric scores (e.g., Mahalanobis distance).
We then set $\tau_{\text{OOD}}$ based on quantiles of these scores, using commonly adopted thresholds such as the 90th, 95th, or 99th percentile, following prior work \cite{liu2020energy}.
Prompts whose OOD scores exceed $\tau_{\text{OOD}}$ are flagged as out-of-distribution.

We evaluate OOD detection performance on our OOD detection testing dataset, containing 1,050 OOD prompts and 954 in-distribution prompts (See Section \ref{sec:dataset_prep}).
We measure performance using precision, recall, and F1-Score, where correctly detected OOD prompts are treated as true positives. Precision measures the proportion of correctly identified OOD prompts among all prompts flagged as OOD, recall measures the proportion of actual OOD prompts that were correctly detected, and F1-Score provides a balanced measure combining both precision and recall.


\subsubsection*{\underline{\textbf{Results}}}

\begin{figure}
    \centering
    \includegraphics[width=1\linewidth]{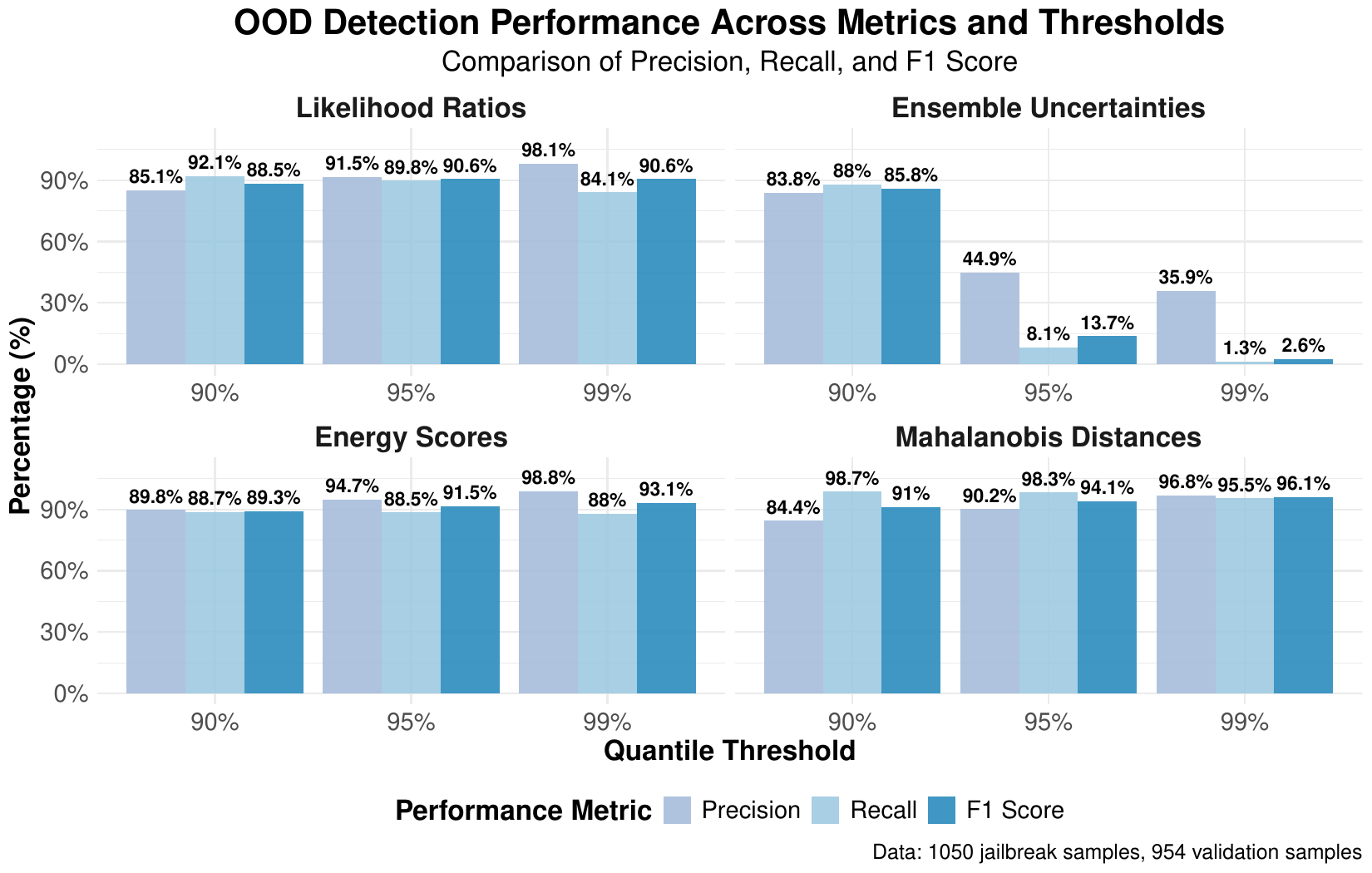}
    \vspace{-0.3cm}
    \caption{(RQ1 Results) OOD detection performance of each OOD method evaluated in RQ1.}
    \label{fig:RQ1_OOD}
\end{figure}


Fig. \ref{fig:RQ1_OOD} presents the Precision, Recall, and F1-Score of the four OOD detection methods, evaluated under three quantile thresholds (90th, 95th, 99th).

\textbf{The Mahalanobis Distances used in our framework achieves the highest F1-Score of 96.1\% when the OOD detection threshold $\tau\_{\text{OOD}}$ is set to the 99th quantile threshold.}
Across all methods, increasing the detection threshold $\tau\_{\text{OOD}}$ generally improves precision while reducing recall. This is because a higher threshold makes the detector more conservative, flagging fewer inputs as OOD, which reduces false positives (improving precision) but may miss some actual OOD samples (reducing recall).
Notably, the Likelihood Ratios and Energy Scores present less balanced performance, with their precision-recall trade-offs resulting in lower F1 scores across all thresholds.
On the other hand, the balanced F1-Score of 96.1\% confirms that Mahalanobis Distances provides the optimal trade-off between precision and recall.
These results confirm the feasibility of leveraging OOD detection to enable guardrails to recognize unseen inputs continually. Among all methods, the Mahalanobis Distances proves to be the most effective in identifying OOD samples while maintaining a low false positive rate.

\subsection{RQ2: \rqtwo}
\subsubsection*{\underline{\textbf{Approach}}}
\begin{figure*}[h]
    \centering
    \includegraphics[width=\textwidth]{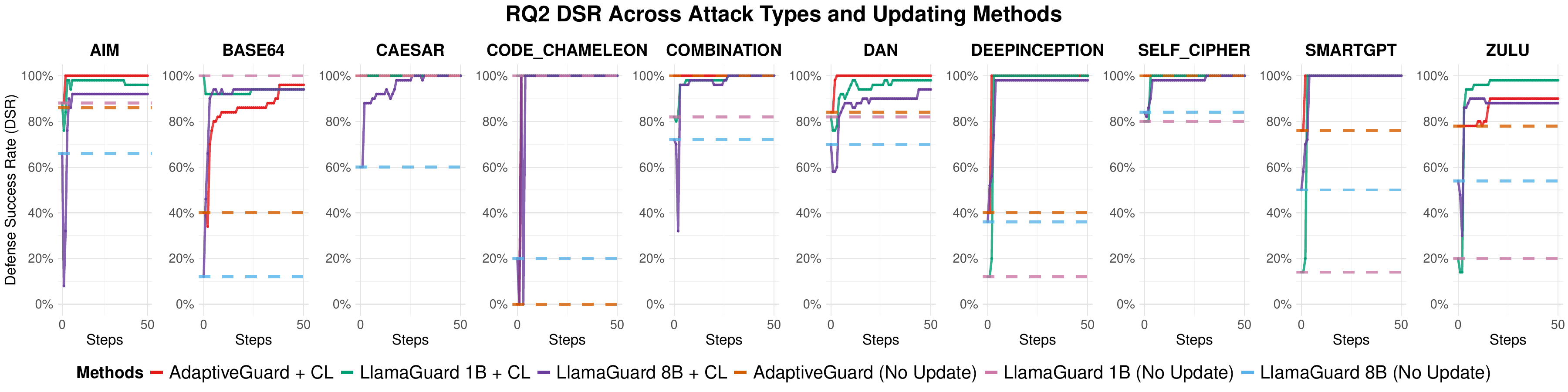}
    \caption{(RQ2 Results) The performance comparison of our \ourapp~and LlamaGuard on OOD prompts when continuously updated through detected OOD prompts.}
    \label{fig:RQ2_DSR}
\end{figure*}
To address this RQ, we compare our \ourapp~approach with LlamaGuard~\cite{inan2023llama}, a state-of-the-art open-source runtime guardrail for LLM-powered systems. Although LlamaGuard shows strong performance on in-distribution data, our analysis reveals its limitations in handling unsafe prompts that differ significantly from its training distribution. We use the LlamaGuard-3-1B and LlamaGuard-3-8B models from Hugging Face as baselines.

To simulate continual learning under OOD detection, we construct a Continual Learning Testing Dataset of 992 OOD jailbreak prompts, all correctly identified by our method in RQ1 (see Section~\ref{sec:dataset_prep}). These prompts span ten types of jailbreak attacks, each exhibiting distinct patterns that bypass guardrails trained on natural unsafe prompts, making them valid OOD scenarios.

In real-world settings, jailbreaks tend to emerge chronologically as adversaries innovate new attack patterns. To reflect this, we organize the dataset into sequential attack waves, where each wave represents a set of related jailbreak techniques. Instead of training on all attacks simultaneously, models are updated incrementally.

For each wave, we hold out 50 samples for testing and use the remaining 50 OOD samples for continual learning. Prompts are fed to the model one at a time. At each time step, we first run inference to log the model’s prediction, then update the model using the same sample. This ensures no data leakage between training and evaluation. The process continues iteratively until all non-hold-out samples are used.

We evaluate two guardrails: our trained \ourapp~from RQ1 and the LlamaGuard baselines.
For each attack wave, all models are initialized from the same pre-trained checkpoint to ensure a fair and consistent comparison.
All models are updated using the same LoRA-based continual learning procedure after running inference.
We measure each guardrail’s effectiveness using the Defense Success Rate (DSR), defined as the number of jailbreak prompts correctly predicted as unsafe, divided by the total number of jailbreak prompts.
This setup allows us to assess how rapidly and effectively each guardrail adapts to unseen OOD jailbreaks using limited adaptation data.

\subsubsection*{\underline{\textbf{Results}}}
Fig. \ref{fig:RQ2_DSR} presents the time-wise DSR across each attack wave, showing how \ourapp, LlamaGuard-1B, and LlamaGuard-8B defend against OOD prompts over time.
We also present three additional baselines where the continual learning (CL) is not applied.
Each subplot represents an attack wave. The x-axis indicates the number of consecutive update steps, while the y-axis shows the DSR on the 50 held-out unsafe samples for each attack.

\textbf{Our~\ourapp~+ Continual Learning (CL), consistently shows the fastest adaptation across most attack waves, achieving optimal DSR within 2 to 38 update steps, with a median of 2 update steps.
In comparison, LlamaGuard requires between 4 to 44 update steps to achieve optimal DSR across the same attack waves, with a median of 4 steps. Guardrails that do not incorporate our continual learning (CL) framework maintain the same level of performance across time steps for each attack wave, as they are not updated with detected OOD prompts.}
\textbf{These results confirm that our CL framework is crucial for enabling adaptive defenses, allowing the guardrails to continually improve and remain effective against evolving and previously unseen jailbreak attacks.}

We also found that different attack patterns exhibited varying levels of complexity for the models to adapt to, significantly impacting their adaptation speed.
In particular, template based attacks such as AIM and SmartGPT converged to optimal DSR in only 2 updating steps for \ourapp~+ CL, and around 4-6 steps for LlamaGuard 8B + CL. 
On the other hand, attacks like Base64 and Zulu presented more significant challenges. Both ~\ourapp + CL and LlamaGuard + CL presented notably slower adaptation speed and often plateaued at optimal DSR values lower than those of less sophisticated attacks.
This indicates that these more complex jailbreak attacks require more extensive adaptation.

\subsection{RQ3: \rqthree}

\begin{figure*}[htbp]
    \centering
    \includegraphics[width=\textwidth]{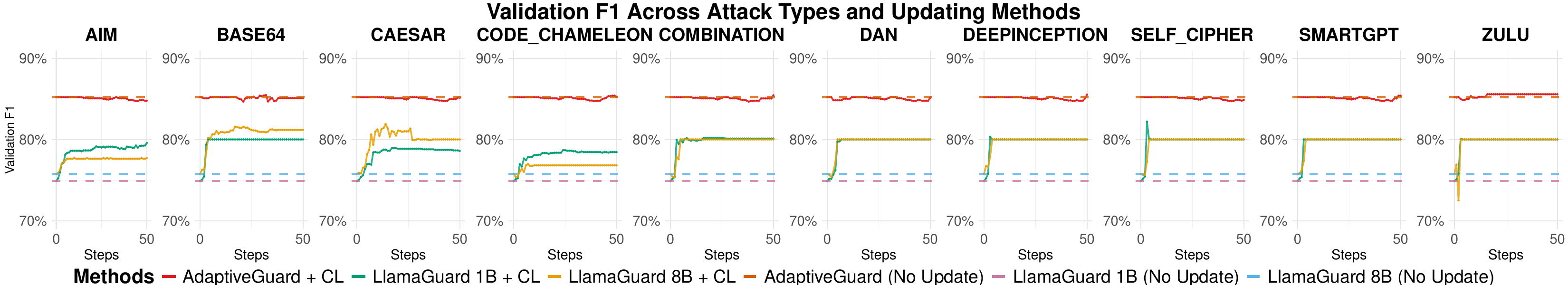}
    \caption{(RQ3 Results) The performance comparison of our \ourapp~and LlamaGuard on in-distribution prompts when continuously updated through detected OOD prompts.}
    \label{fig:RQ3_F1}
\end{figure*}

\subsubsection*{\underline{\textbf{Approach}}}
To answer this RQ, we reuse the experimental setup from RQ2 for continual learning (CL), with one key difference: after each continual update step, we evaluate on in-distribution (ID) data instead of out-of-distribution (OOD). This allows us to assess knowledge retention of ID prompts after updating with each OOD prompt. Specifically, we use the Aegis validation set—part of the dataset from RQ1—containing 314 safe and 640 unsafe natural language prompts. We measure performance using the F1-score to account for both safe and unsafe classes.




\subsubsection*{\underline{\textbf{Results}}}
Fig.~\ref{fig:RQ3_F1} presents the time-wise F1-Score across each attack wave, showing how \ourapp, LlamaGuard-1B, and LlamaGuard-8B defend against in-distribution prompts over time when continuously updated with OOD prompts.
Similar to RQ2, we present three additional baselines where the continual learning (CL) is not applied.
Each subplot represents an attack wave. The x-axis indicates the number of consecutive update steps, while the y-axis shows the F1-Score on the 954 in-distribution prompts from the Aegis validation set.

\textbf{\ourapp~+CL achieves the highest median F1-Score of 85\% on in-distribution prompts across all attacks after the final update with OOD prompts, with the F1-Score ranging from 85\% to 86\%.}
In comparison, the best-performing baseline, LlamaGuard-8B, achieves a median F1-Score of 80\% on in-distribution prompts, with a minimum of 77\% and a maximum of 81\%.

Across all attack types, \ourapp~+ CL consistently maintains the highest in-distribution F1-Score throughout the CL process, with only slight variation (ranging from –0.4\% to +0.4\%).
This indicates that \ourapp~+CL is able to learn to defend against unseen jailbreak attacks without sacrificing its learned in-distribution knowledge.
\textbf{In summary, \ourapp~+ CL maintains the highest performance on in-distribution prompts compared to all baselines, demonstrating minimal forgetting even after continuous updates with detected OOD prompts.}

\section{Discussion}
\label{sec:discussion}
In the previous section, we evaluated the effectiveness of our proposed \ourapp~approach for OOD detection and continual learning against other baselines.
While our results demonstrate clear advancements, it remains unclear (1) whether our approach can still perform well in a setting where a single continual model is built across all attacks, rather than reinitialising the model for each attack; (2) what the trade-offs are between different continual learning (CL) update methods; and (3) what is the computational efficiency of our approach compared to other baselines. Thus, we evaluate a single continual model across all attacks, compare LoRA with full SFT as update strategies, and analyze the computational efficiency of each approach.

\begin{table}[h]
\centering
\caption{(Discussion) DSR of \ourapp~under Sequential Continual Learning with and without Early Stopping across 10 OOD Attack Types.}
\label{tab:lora_dsr_results}
\resizebox{\linewidth}{!}{%
\begin{tabular}{lcc}
\toprule
\textbf{Attack Type} & \textbf{DSR (No Early Stopping)} & \textbf{DSR (Early Stopping)} \\
\midrule
AIM Attack & 1.00 & 1.00 \\
Base64 Attack & 1.00 & 0.86 \\
Caesar Attack & 1.00 & 1.00 \\
Code Chameleon Attack & 1.00 & 1.00 \\
Combination Attack & 1.00 & 1.00 \\
DAN Attack & 1.00 & 1.00 \\
DeepInception Attack & 1.00 & 1.00 \\
Self Cipher Attack & 1.00 & 1.00 \\
SmartGPT Attack & 1.00 & 1.00 \\
Zulu Attack & 1.00 & 0.92 \\
\midrule
\textbf{Average} & \textbf{1.00} & \textbf{0.98} \\
\bottomrule
\end{tabular}
}
\end{table}

\subsection{Continual Performance Without Model Reinitialization}
In a typical industry setup, a guardrail is deployed as an additional layer before user inputs reach the underlying LLM. This guardrail often operates as a single continually updated model that incrementally incorporates new external knowledge—such as OOD prompts—to adapt over time. To reflect this real-world scenario, we construct a single continual version of \ourapp, which is updated sequentially across all 10 attacks without reinitialization between them.

We apply continual learning by updating the same \ourapp~model using OOD prompts from each attack in the same experimental dataset used in RQ2. After each update, we evaluate the model's DSR on the test set corresponding to the respective attack. We consider two update settings: in the early stop setting, the continual learning process halts once the model achieves 95\% DSR on the current attack; in the no early stop setting, training continues through all OOD prompts regardless of intermediate performance.

Table~\ref{tab:lora_dsr_results} presents the DSR results for our sequential continual learning setup. The results show that \ourapp~achieves a 100\% DSR across all 10 attacks when trained without early stopping. Even with early stopping enabled, the model maintains an average DSR of 98\%.
Notably, the Obfuscation-Based attacks—Base64 (86\%) and Zulu (92\%)—prove the most challenging to defend under early stopping. This suggests that these attacks require more adaptation steps for the model to fully internalize their patterns in a guardrail.
Overall, these results indicate that our continual learning framework not only adapts effectively to new attack types but also retains knowledge of previously encountered ones. This makes it well-suited for real-world deployment, where a single guardrail must continuously evolve to defend against emerging threats without forgetting past vulnerabilities.

\begin{table}[htbp]
\centering
\caption{(Discussion) Comparison of SFT vs LoRA Performance on Jailbreak Defense}
\label{tab:sft_lora_comparison}
\begin{tabular}{lcccc}
\toprule
\multirow{2}{*}{\textbf{Statistic}} & \multicolumn{2}{c}{\textbf{Full Fine-Tuning}} & \multicolumn{2}{c}{\textbf{LoRA}} \\
\cmidrule(lr){2-3} \cmidrule(lr){4-5}
& \textbf{DSR} & \textbf{F1-Score} & \textbf{DSR} & \textbf{F1-Score} \\
\midrule
Min & 1.00 & 0.800 & 0.90 & 0.848 \\
Median & 1.00 & 0.800 & 1.00 & 0.852 \\
Max & 1.00 & 0.801 & 1.00 & 0.856 \\
\midrule
\textbf{Average} & \textbf{1.00} & \textbf{0.800} & \textbf{0.99} & \textbf{0.852} \\
\bottomrule

\end{tabular}
\end{table}

\subsection{Trade-offs Between LoRA and Full Fine-Tuning for CL}
In our continual learning (CL) framework, we adopted LoRA-based adaptation for updating the guardrails.
LoRA is known for its computational efficiency, as it updates only a small subset of parameters while keeping the rest of the pre-trained model frozen. This selective adaptation can lead to better knowledge retention—reflected as stronger in-distribution performance in our context—and has been shown to mitigate catastrophic forgetting compared to full-parameter fine-tuning~\cite{hu2022lora,he2022towards}. While our results confirm LoRA's effectiveness, it remains unclear whether it consistently offers better knowledge retention than full fine-tuning in the context.


To investigate this, we compare our LoRA-based approach~\cite{hu2022lora} with full fine-tuning~\cite{howard2018universal} using the same experimental setup as in RQ2 and RQ3. Since the evaluation spans ten different attacks, we report the minimum, median, and maximum values for both DSR (representing performance for OOD prompts) and F1-Score (performance for in-distribution prompts).
Table~\ref{tab:sft_lora_comparison} presents the performance comparison between LoRA and full fine-tuning across all 10 attacks. Our LoRA-based approach achieves a median DSR of 100\% and a median F1-Score of 85\% on in-distribution prompts, indicating strong knowledge retention. In contrast, full fine-tuning also achieves a median DSR of 100\% but yields a lower median F1-Score of 80\%. These results are consistent with prior findings in the continual learning literature and demonstrate that our LoRA-based method more effectively preserves previously learned knowledge while adapting to new threats.


\begin{table}[h]
\centering
\vspace*{-0.5em}
\caption{(Discussion) Training \& inference time per sample, and memory usage during training}
\label{tab:computation_requirements}
\setlength{\tabcolsep}{4pt}
\begin{tabular}{lccc}
\toprule
\textbf{Metric} & \textbf{AdaptiveGuard} & \textbf{LlamaGuard 1B} & \textbf{LlamaGuard 8B} \\
 & \textbf{LoRA} & \textbf{LoRA} & \textbf{LoRA} \\
\midrule
Training Time & 0.60s & 1.06s & 2.04s \\
Inference Time & 0.01s & 0.25s & 1.10s \\
Memory Usage & 1.3 GB & 4.0 GB & 27.1 GB \\
\bottomrule
\vspace*{-1.5em}
\end{tabular}
\end{table} 

\subsection{\hbox{Computational Efficiency: \ourapp~and Baselines}}
In the context of practical deployment of our \ourapp, computational resources are a key consideration factor.
Therefore, we analyse (1)~the training time per iteration (update step), (2)~inference time per testing sample, and (3)~training and testing memory usage.
We conduct the analysis using the datasets from RQ2 and RQ3, comprising 992 jailbreak attacks, 314 safe, and 640 unsafe NL prompts.
We compare our approach (\ourapp+LoRA) with the baselines LlamaGuard-1B+LoRA and LlamaGuard-8B+LoRA.

Table \ref{tab:computation_requirements} summarises the computational efficiency of our approach compared to baseline methods during both training and inference.
Compared to the LlamaGuard baselines, our method demonstrates substantially greater efficiency.
Specifically, compared to LlamaGuard-1B and LlamaGuard-8B, \ourapp~achieves 43\% and 71\% faster training times during continual learning (CL), delivers 25× and 110× faster inference, and reduces memory usage by 67\% and 95\%, respectively.
These substantial gains stem from the compact architecture of our base model, GPT-2 (approximately 137M parameters), in contrast to the considerably larger LlamaGuard models with 1B and 8B parameters.
These results highlight the practicality of our approach for resource-constrained settings.

\section{Threats to Validity}
\label{sec:threats}
\textbf{Threats to construct validity} relate to the selection of jailbreak attacks and OOD thresholds.
We selected 10 different jailbreak attacks guided by their prevalence to guardrails and ability to represent a wide spectrum of jailbreak techniques \cite{dong2024safeguarding, yi2024jailbreak, wei2024jailbroken}. 
While additional attack types could be included in future evaluations, this would not fundamentally alter the key conclusions presented in our research questions: that complex jailbreak attacks can be effectively detected by our proposed OOD detection measures (RQ1), and that detected attacks can be efficiently learned by \ourapp through continual learning with minimal catastrophic forgetting (RQ2 and RQ3).
For the OOD threshold, we adopted commonly used quantile-based thresholds (90\%, 95\%, and 99\%) based on the distribution of OOD scores on the training set~\cite{liu2020energy,lee2018simple}. This approach allows us to select the most appropriate threshold values based on our experiments, ensuring that only the most anomalous prompts are flagged for guardrail adaptation. However, in practice, threshold selection is based on various factors depending on the application's tolerance for risk and the distributional properties of the data. 

\textbf{Threats to internal validity} relate to the potential influence of hyperparameter settings during the fine-tuning of our \ourapp~and the selection of OOD detection thresholds. Variations in energy threshold values, LoRA configurations, or learning rates, compared to those specified in Section \ref{sec:exp_design}, could impact experiment's outcomes. 
To address these threats, we open-source our replication package and provide detailed documentation of all hyperparameter settings to ensure the experiment is reproducible by future researchers. 

\textbf{Threats to external validity} concern the generalizability of our results. Our experiment findings are supported by the dataset, jailbreak methods, and model architecture employed during the study. Our training dataset contains 12,233 unsafe prompts and 6,905 safe prompts. Our  validation dataset contains 314 safe and 640 unsafe natural language prompts to assess false positive rates and catastrophic forgetting of~\ourapp. We also applied the 10 jailbreak attacks to 105 separate unsafe prompts, resulting in 1,050 jailbreak prompts across 10 categories for RQ1. Of the 1,050 jailbreak prompts, 992 are identified as OOD and used to study the continual learning adaptation in RQ2 and RQ3.
Our evaluation focuses on GPT-2 as the base model, which represents a widely-studied language model in AI safety research, but may not capture the behaviors of larger or more recent language models.
While \ourapp~is fine-tuned specifically to address the jailbreak attacks and model configurations discussed in this paper, other prompt datasets, jailbreak methods, and language model architectures can be explored in future work.


\section{Conclusion}
In this paper, we first show that state-of-the-art guardrails like LlamaGuard face a critical limitation, with their Defense Success Rate (DSR) dropping to as low as 12\% against jailbreak attacks not seen during training.
To address this, we present \ourapp, an OOD-aware continual learning framework that detects and adapts to previously unseen jailbreak patterns by treating them as out-of-distribution (OOD) inputs. This builds on the observation that guardrails like LlamaGuard are trained on NL unsafe prompts, while jailbreaks often use obfuscated inputs that fall outside their training distribution.
To evaluate \ourapp, we compile a dataset using ten state-of-the-art jailbreak methods: AIM, DAN, Self Cipher, Deep Inception, SmartGPT, and Code Chameleon.
Through our evaluation, we found that \ourapp~achieves a 96\% true positive rate in OOD detection and reaches 100\% DSR within a median of two update steps—twice as fast as LlamaGuard under the same continual learning setup.
Moreover, \ourapp~retains 85\% F1-score on in-distribution data after adaptation, outperforming LlamaGuard’s 80\%. These results demonstrate that \ourapp~could offer an effective post-deployment solution for adaptive jailbreak defense in dynamic production.


\balance
\bibliographystyle{IEEEtran}
\bibliography{reference}

\end{document}